# Electronic quantum effect on hydrogen bond geometry in water dimer

Danhui Li[1,2], Zhiyuan Zhang[1,2] Wanrun Jiang[1,2] Depeng Zhang[1,2] Yu Zhu[1,2] and Zhigang Wang[1,2*]

[1]Institute of Atomic and Molecular Physics, Jilin University, Changchun 130012, China.

[2]Jilin Provincial Key Laboratory of Applied Atomic and Molecular Spectroscopy (Jilin University), Changchun 130012, China.

[*]Corresponding author. E-mail: wangzg@jlu.edu.cn.

## Abstract

During compression of a water dimer calculated with high-precision first-principles methods, the trends of H-bond and O-H bond lengths show quantum effect of the electronic structure. We found that the H-bond length keeps decreasing, while the O-H bond length increases up to the stable point and decreases beyond it when the water dimer is further compressed. The remarkable properties are different from those observed in most previous researches which can be understood and extrapolated through classical simulation. The observations can be explained by the decrease in orbital overlap and change in the exchange repulsion interaction between water monomers. The dominant interaction between water monomers changes from electrostatic interaction to exchange repulsion at the turning point of the O-H bond length when the O···O distance is decreased. These findings highlight the quantum effect on the hydrogen bond in extreme conditions and play an important role in the recognition of the hydrogen bond structure and mechanism.

## Introduction

Water covers two thirds of our planet, and is the source and central part of all life forms. Water is anomalous and strange in that it expands on freezing and is the densest in the liquid state at 4℃ rather than becoming steadily denser with cooling. In addition, it has an abnormally high heat capacity, odd viscosity, and more interesting properties[1]. Most of these anomalies are rationalized by the standard tetrahedral hydrogen-bonded network[2,3]. Besides, numerous studies have been carried out on water through first-principles methods by extrapolating from the classical model[4,5]. However, it was not until the discovery of nuclear quantum effect of the hydrogen bond that the concepts from classical simulation were refreshed[6,7]. The quantized behavior of electrons[8] and the reported electron delocalization[9] inspired us to study the electronic quantum effect of the hydrogen bond in water, which has not been found in previous researches.

## Results and Discussion

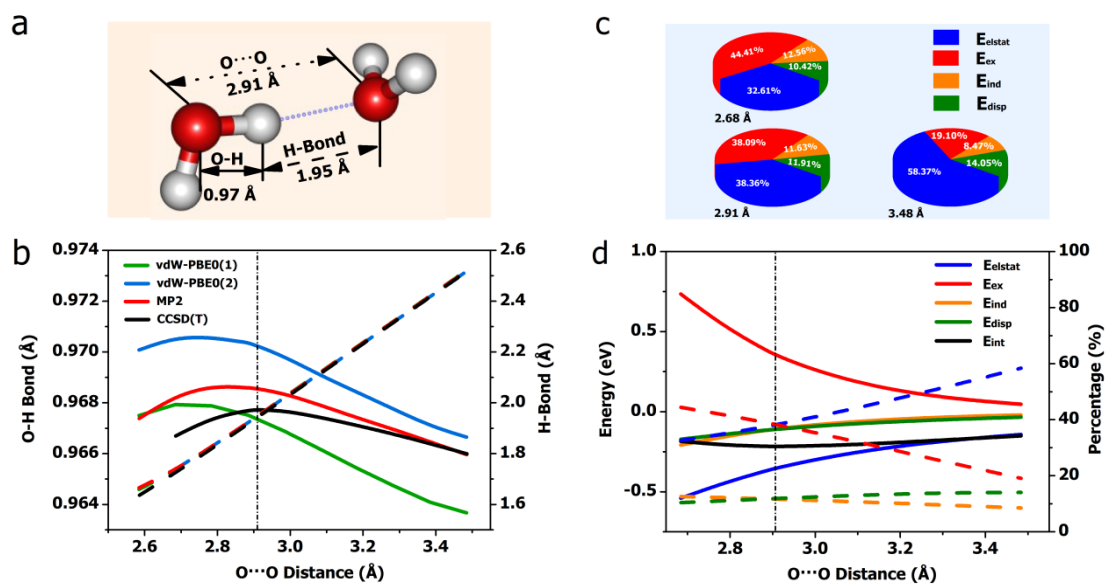

**Figure 1| Electronic quantum effect on the H-bond and O-H bond lengths in the water dimer.**

**(a)** The stable structure optimized with CCSD(T) method. The O⋯O distance, O-H bond length, and H-bond length are 2.91, 0.97, and 1.95 Å, respectively. **(b)** Change in O-H bond length (H-bond length) with change in O···O distance in $(H_2O)_2$, which is represented with a solid line (dash line) for each high-precision first-principles method. vdW-PBE0(1) and vdW-PBE0(2) denote the PBE0-D3 method with aug-cc-pVTZ and the same method with slater TZP basis set. The black dotted line corresponds to the stable structure optimized with the CCSD(T) method with an O···O distance of 2.91 Å. **(c)** Percentages of the four interaction terms in the total interaction energy at O···O distances of 2.68, 2.91 and 3.48 Å. $E_{elstat}$, $E_{ex}$, $E_{ind}$, and $E_{disp}$ represent electrostatic interaction, exchange repulsion, induction interaction, and dispersion, respectively. **(d)** Energy decomposition are applied in the water dimer structures with CCSD(T) method and aug-cc-pVTZ basis set, the black dotted line denotes the turning point of 2.91 Å, where the dominant interaction between the water monomers changes from electrostatic attraction to exchange repulsion.

In this work, the unique quantum effect of the electronic structure of water was determined by simulated scanning. Different high-precision first principles methods including singles and doubles coupled cluster methods with perturbative triple excitations (CCSD(T)), second-order Møller-Plesset perturbation theory (MP2), and density functional theory (DFT) were applied. The stable structure optimized with CCSD(T) is shown in Fig. 1a. From the results of different methods, it was concluded that when the water dimer is compressed, the H-bond length continuously decreases, while the O-H bond length increases up to the stable point, beyond which it decreases. The trends of change in H-bond length and O-H bond length are shown in Fig.1b (see SI for details). This is different from the previous compression results of water[10,11] and cannot be implemented in classical simulation because of the exchange repulsion of electrons in the quantum mechanics specially.

To explain the mechanism and the contribution of the exchange repulsion further, we carried out Symmetry-Adapted Perturbation Theory (SAPT)[12] based on the quantum mechanics and it can be applied on the interaction of two water monomers, it decomposes the interaction into physical components of electrostatic interaction, exchange repulsion, induction, and dispersion. The percentages of the four terms in the total interaction and the corresponding energies are provided in Fig. 1c and Fig. 1d, respectively. For the decrease in H-bond length with compression of water dimer, it is related to exchange repulsion, resulting from orbital overlap. Because of the decrease in O···O distance, the orbital overlap of the water monomers increases, leading to an increase in the exchange repulsion and decrease in bond length of the H-bond connecting the monomers. The relationship between bond length and orbital overlap as well as exchange repulsion is consistent with the previous report[13]. For the O-H bond length increases up to the stable point, beyond which it decreases. We concluded that when the O···O distance in the water dimer structures is larger than that in the stable structure, the electrostatic interaction significantly contributed to the total interaction between the water monomers. Within this region, attraction from the electrostatic term becomes larger with compression, leading to an increase in the O-H bond length. Contrarily, when the O···O distance in the scanned water dimer structures is smaller than that in the stable structure, the total interaction is dominated by the larger exchange repulsion, leading to a decrease in the O-H bond length (see SI for details). The physical basis of the exchange repulsion between water monomers is a pure quantum effect, which changes the parameters of the water monomers including the O-H bond length[14].

In summary, this work focuses on the quantum effect of the electronic structure on the changes in the O-H bond length and the H-bond length with a decrease in the O···O distance in a water dimer, which is of significance for the related study of quantum effect in hydrogen bond systems in the future.

**Methods**

The structures of the $(H_2O)_2$ are scanned with high precision first principles methods including singles and doubles coupled cluster methods with perturbative triple excitations (CCSD(T)), second-order Møller-Plesset perturbation theory (MP2), and PEB0-D3 method of density functional theory (DFT) with aug-cc-pVTZ basis set in Gaussian 09[15] and PBE0-D3 method with aug-cc-pVTZ basis set in Amsterdam Density Functional package[16-18] respectively, the comparison of the results about the bond length, optimized energy, variation of the bond angle are provided, which proves all of the accuracy of the results.

The energy decomposition of the $(H_2O)_2$ scanned with CCSD(T) is implemented in the Psi4[19,20], the interaction energies of the water monomers is given by $E_{int}=E_{elstat}+E_{ex}+E_{ind}+E_{disp}$, which divided the interaction energy of the water monomers into four parts: the electrostatic term, the exchange repulsion term, the induction term and the dispersion term and the sum of the electrostatic term and the exchange repulsion term is related to the variation of the conformation of $(H_2O)_2$ closely.

Electrostatic interaction includes Coulombic multipole-multipole-type interactions as well as the interpenetration of charge clouds. Exchange repulsion is a repulsive force that arises from monomer wavefunction overlap and the fermionic anti-symmetry requirements of the dimer wavefunction. Induction includes polarization from each monomer's response to the other's electric field as well as charge transfer. Dispersion is an attractive force resulting from the dynamic correction between the electrons on one monomer with those on another.

## Acknowledgments

This work was supported by the National Science Foundation of China (under grant number 11374004 and 11674123). We also acknowledge the assistance of the High Performance Computing Center of Jilin University.

## Competing Interests

The authors declare no competing interests.

## Supporting Information

Figures S1-S2

Tables S1-S7